\documentclass[
reprint,
amsmath,
amssymb,
aps,
superscriptaddress,
]{revtex4-2}
\usepackage[colorlinks=true,allcolors=blue]{hyperref}
\usepackage{graphicx,color}
\usepackage{dcolumn}
\usepackage{bm}
\usepackage{subfigure}
\usepackage{amsmath,amsfonts,amssymb}
\usepackage{acronym}
\usepackage{enumitem}
\usepackage{array}
\usepackage{multirow}
\usepackage{soul}

\allowdisplaybreaks
\newcommand{\TRC}{MOE Key Laboratory of TianQin Mission,
TianQin Research Center for Gravitational Physics $\&$ School of Physics and Astronomy,
Frontiers Science Center for TianQin,
Gravitational Wave Research Center of CNSA,
Sun Yat-sen University (Zhuhai Campus), Zhuhai 519082, China}
\newcommand{\TRCLu}{Southwestern Institute of Physics, Chengdu 610041, China}
\newcommand{\TRCTang}{State Key Laboratory of Space Weather, National Space Science Center, \\Chinese Academy of Sciences, Beijing 100190, China}

\begin{document}

\title{Plasma noise in TianQin time delay interferometry}

\author{Yi-De Jing}
\affiliation{\TRC}
\author{Lu Zheng}
\affiliation{\TRC}
\author{Shutao Yang}
\affiliation{\TRC}
\author{Xuefeng Zhang}
\email{zhangxf38@sysu.edu.cn}
\affiliation{\TRC}
\author{Lingfeng Lu}
\email{lulingfeng@swip.ac.cn}
\affiliation{\TRCLu}
\author{Binbin Tang}
\affiliation{\TRCTang}
\author{Wei Su}
\affiliation{\TRC}
\date{\today}

\begin{abstract}
TianQin is a proposed geocentric space-based gravitational wave observatory mission, which requires time-delay interferometry (TDI) to cancel laser frequency noise. With high demands for precision, solar-wind plasma environment at $\sim 10^5$ km above the Earth may constitute a non-negligible noise source to laser interferometric measurements between satellites, as charged particles perturb the refractivity along light paths. In this paper, we first assess the plasma noises along single links from space-weather models and numerical orbits, and analyze the time and frequency domain characteristics. Particularly, to capture the plasma noise in the entire measurement band of $10^{-4} - 1$ Hz, we have performed additional space-weather magnetohydrodynamic simulations in finer spatial and temporal resolutions and utilized Kolmogorov spectra in high-frequency data generation. Then we evaluate the residual plasma noises of the first- and second-generation TDI combinations. Both analytical and numerical estimations have shown that under normal solar conditions the plasma noise after TDI is less than the secondary noise requirement. Moreover, TDI is shown to exhibit moderate suppression on the plasma noise below $\sim 10^{-2}$ Hz due to noise correlation between different arms, when compared with the secondary noise before and after TDI. 
\end{abstract}
\maketitle


\section{Introduction}

Till now, more than 90 gravitational wave (GW) events have been detected by ground-based GW detectors \cite{LIGOGW150914,LIGOGW151226,LIGOimage,LIGOScientific2021djp,LIGOScientific:2021usb,LIGOScientific:2020ibl}. However, due to relatively short arm-lengths and seismic noises, ground-based detectors so far are limited to detecting GW signals at high frequencies $f\gtrsim10~\rm{Hz}$. Complementary to ground-based detectors, space-based GW detectors, such as LISA \cite{LISA2017arxiv}, TianQin \cite{Luo2016}, Taiji \cite{HuWR2017Taiji}, B-DECIGO \cite{Kawamura2018DECIGO}, etc., will allow GW detection in the millihertz frequency band.

TianQin is a proposed space-based GW observatory mission consisting of three identical satellites in drag-free motion around the Earth. The orbital radius of TianQin is $\sim1\times10^{5}~\rm{km}$, and the normal of the orbital plane points to the reference source RX J0806.3+1527 \cite{Luo2016}. The working principle of TianQin is to measure the relative distance changes caused by GWs between free-falling test masses inside different satellites using inter-satellite laser heterodyne interferometry. 

Unlike ground-based detectors, orbital perturbations to the satellites will cause unequal and time-varying arm-lengths in space-based detectors (e.g., \cite{YeBB2019}). The unequal arm-lengths couple with fluctuations of laser frequencies and result in a troublesome laser frequency noise in science measurements. This noise is normally 6-9 orders of magnitude greater than the sensitivity requirements. The problem can be solved by the time-delay interferometry (TDI) technique \cite{Tinto2002de,Armstrong2003ut,Tinto2014lxa}. TDI allows suppression of laser frequency noise by linear combinations of instantaneous and time-delayed one-way phase measurements. The underlying principle is to reduce the unequal arm-lengths by composing a virtual nearly equal arm-length interferometer in data processing to achieve common-mode rejection of laser frequency noise. The first-generation TDI combinations was initially proposed for a static detector configuration \cite{Tinto2014lxa}. But if one considers actual TianQin orbits, the first-generation TDI combinations may not satisfy the requirement at certain frequencies. Hence we also need to take into account the second-generation TDI combinations to further reduce the unequal arm-lengths \cite{Zhou2021PRD,Shaddock2003dj}. Apart from laser frequency noise, the extended TDI combinations can also suppress clock noise \cite{Otto2012dk,Tinto2018PRD,Hartwig2020PRD} and optical bench displacement noise \cite{Otto2012dk}. Since TDI forms a crucial part in current designs of space-based GW detection, it would be beneficial to discuss various environmental noises in the context of TDI so as to identify their end effects on data output. Some recent studies on TDI can be found, e.g., in Refs.~\cite{Bayle2018hnm,Bayle2021mue,Tinto2021cwi,Page2021asu,Muratore2021uqj}. 

Recently, Ref.~\cite{LuLF2021raf} has calculated the optical path length fluctuations caused by the solar wind plasma along single arms. The results indicate that the impact is close to the displacement noise requirement of TianQin and constitutes a noticeable noise source. Moreover, the effect becomes larger when a strong magnetic storm occurs. In the following discussions, we refer to this type of noise as plasma noise. Subsequently, the work of \cite{SuWei2021zmj} further has evaluated the residual plasma noise in TDI combinations. However, the amplitude spectral densities (ASD) of plasma noise is restricted to $<8~\rm{mHz}$ due to temporal and spatial resolution limitations. Furthermore, only linear interpolation of plasma noise was considered in the first-generation TDI simulations. In addition, in Refs.~\cite{Lu2020wavefrontdistortion,Su2020acceleration}, the effects of space environment on wavefront distortion of inter-satellite laser beams and the acceleration noise of test masses were investigated, respectively, and both are manageable for TianQin. 

For LISA, Refs.~\cite{Smetana2020raf,Jennrich2021okh} have assessed the impact of solar wind plasma on LISA inter-satellite scientific measurements using a statistical model of turbulence and real solar wind observations. The results have show that normal solar activities does not affect LISA in the measurement frequency band. The expected spectral index of the electron number density fluctuation has a value of $\alpha=-1.75\pm0.5$, which is close to the Kolmogorov $f^{-5/3}$ spectrum \cite{Thorne2017,Schmidt2010}. This means that both the fluctuations of the electron number density and the plasma noise of a single link become smaller as the Fourier frequency increases. Besides, the ASD of the magnetic field fluctuations measured by the LISA Pathfinder's magnetometers agrees with the results from the Advanced Compton Explorer (ACE) in the frequency band below $3~\rm{mHz}$. The spectral index is consistent with the Kolmogorov's turbulence spectrum \cite{Armano2020LISAPFMagnetic}.

Similar to the solar wind plasma is the ionosphere that consists of ionized Earth's upper atmosphere by solar radiation. Since both are dispersive media, the electromagnetic wave signal propagates in this media will cause frequency-dependent signal group delay and carrier phase advance with the same magnitude and opposite signs. Using this property, ionospheric effects can be calculated or corrected using multiple operating frequencies. For example, each GPS satellite uses three frequencies to correct the ionospheric effect so as to improve positioning accuracy \cite{Thomas1999}. The GRACE mission uses a dual-frequency microwave (K and Ka bands) ranging system to correct ionospheric delays to achieve micrometer precision \cite{Kim2000PHD}. 

In future space-based GW detection missions with increasing accuracy requirements, plasma noise will become a non-negligible factor. For example, the second-generation space GW detection mission BBO is planned to measure the solar wind plasma noise by adding an auxiliary microwave inter-satellite links \cite{Harry2006CQGBBO,Crowder2005PRDBBO}. For other methods, we previously analyzed a two-color TDI technique \cite{ZhangDTDI}. The core procedure is the addition of $532~\rm{nm}$ laser by frequency doubling technique. However, this increases the system complexity and the technical feasibility needs further assessment. In this paper, we instead focus on how the plasma noise is propagated though TDI data processing and evaluate the suppressing effect. 

As discussed above, high-frequency data was not available to fully analyze the impact of the solar wind plasma on the inter-satellite measurements of TianQin. Also note that the simulation results in Ref.~\cite{SuWei2021zmj} were not explained thoroughly by theory, and preferably, more realistic simulations would be needed to verify these results. Thereby, this work first analyzes the time domain and frequency domain characteristics of plasma noise of a single link, and then, calculates the TDI-combined plasma noise, and compares the theoretical results with the numerical simulations. 

The remainder of this paper is arranged as follows. In Section~\ref{methodology} and \ref{SimulationSetup}, we describe how to calculate the optical path length fluctuations caused by the solar wind plasma on a single arm, including plasma refractive index, MHD simulation and Kolmogorov spectrum. In Section~\ref{Theoretical}, the residual plasma noises after the first- and second-generation TDI combinations are presented analytically, and the simulation results are displayed in Section~\ref{result}. The conclusions are given in  Section~\ref{summary}. 


\section{Refractive index of solar-wind plasma}\label{methodology}

As is well known, the phase refractive index $n$ of space plasma media is given by the Appleton-Hartree (AH) equation. In most practical cases, the AH equation can be simplified to \cite{Petrie2011}
\begin{equation}\label{AHEquation}
  n=1-\frac{X}{2}-\frac{XY}{2}\cos\theta - \frac{X^{2}}{8},
\end{equation}
in which, the quantities $X$ and $Y$ are read off as $ X= \frac{\omega^{2}_{p}}{\omega^{2}}$, $ Y= \frac{\omega_{h}}{\omega}$, respectively, and $ \omega_{p} = \sqrt{\frac{N_{e}e^{2}}{\varepsilon_{0}m}}$, $ \omega_{h}=\frac{B_{0}\left|e\right|}{m}$, $\theta$ is the angle between the propagation direction of the electromagnetic wave and the magnetic field, $N_{e}$ is the  electron number density of solar wind plasma,  $e$ and $m$ represent the charge and mass of the electron, $B_{0}$ is the space magnetic field, $\varepsilon_{0}$ is the vacuum electric permittivity, $\omega$ is the angular frequency of the electromagnetic wave, $\omega_{p}$ and $\omega_{h}$ are the plasma frequency and the electron gyro-frequency.

The solar wind plasma is mainly composed of electrons, protons, and energetic particles from the Sun. Since electrons are much lighter than protons thus respond much faster than the oscillating fields. As a result, we only consider the electrons' contributions. Therefore, Eq. (\ref{AHEquation}) can be written equivalently as \cite{Petrie2011}
\begin{equation}
n=1-\frac{1}{2}\frac{A_{p}N_{e}}{f^{2}}-\frac{A_{p}N_{e}A_{g}B_{0}\cos\theta}
{2f^{3}}-\frac{1}{8}\frac{A^{2}_{p}N^{2}_{e}}{f^{4}},
\end{equation}
where $A_{p}=\frac{e^2}{4\pi^{2}\varepsilon_{0}m}=80.62 ~\rm m^{3}\rm s^{-2}$, and $A_{g}=2.80\times10^{10}~\rm{sA/kg}$. In this simplified expression, the term proportional to $f^{-2}$ becomes the first order term. The second and third order terms are inversely proportional to the third and fourth powers of $f$.

For TianQin, the frequency of the laser is $f\approx 2.8\times 10^{14}~\rm{Hz}$ ($\lambda=1064~\rm{nm}$), so the main contribution to the phase refractive index comes from the first order term ($\frac{X}{2}$) \cite{SuWei2021zmj}, and we only consider the lowest order term in the following discussions. Substituting these values into Eq. (\ref{AHEquation}), we further obtain
\begin{equation}
n\approx1-\frac{X}{2}=1-\frac{1}{2}\frac{A_{p}N_{e}}{f^{2}}=
1-40.3\frac{N_{e}}{f^{2}}.
\end{equation}
Consequently, the signal delay caused by the solar wind plasma can be expressed as
\begin{equation}\label{signaldelay}
s_{ij}(t)=\frac{40.3}{f^{2}}\int_{i}^{j} N_{e}\mathrm{d}l=\frac{40.3}{f^{2}}D_{\rm{TEC}},
\end{equation}
where $D_{\rm{TEC}}$ stands for the total electron content (TEC) along the laser path from spacecraft $i$ to spacecraft $j$.

Eq. (\ref{signaldelay}) shows that the effect of plasma on laser interferometric measurement depends on the integral electron number density fluctuations along the laser propagation path. For typical values of the involved quantities, i.e. $N_{e} \sim 10 ~\rm cm^{-3}$ at an altitude of $\sim 10^{5}~\rm{km}$ and the arm-length $1.7\times10^{5}~\rm{km}$, we arrive at $s_{ij} \approx 1$ pm \cite{LuLF2021raf}. This motivates us to look further into the frequency domain behavior, and the single-link displacement noise requirement (preliminary) for plasma is given by
\begin{equation} \label{eq_req}
S^{1/2}_{\rm pla}\le 0.3\, \frac{\rm pm}{\sqrt{\rm Hz}} \sqrt{1+\left(\frac{7\rm mHz}{f}\right)^{4}}~, \ 0.1{\rm mHz}\leq f \leq 1{\rm Hz}, 
\end{equation}
which is allocated from the total noise requirement \cite{Luo2016,Mei2020}. Here $7$ mHz is the corner frequency where the acceleration noise and the interferometer measurement noise intersect in the sensitivity curve. In the following, we will discuss the plasma displacement noise along a single arm. 

\section{Single-link plasma noise}\label{SimulationSetup}

\subsection{MHD simulations}\label{MHD_simulation}

\begin{figure}[ht]
\includegraphics[width=0.5\textwidth, height=0.4\textwidth]{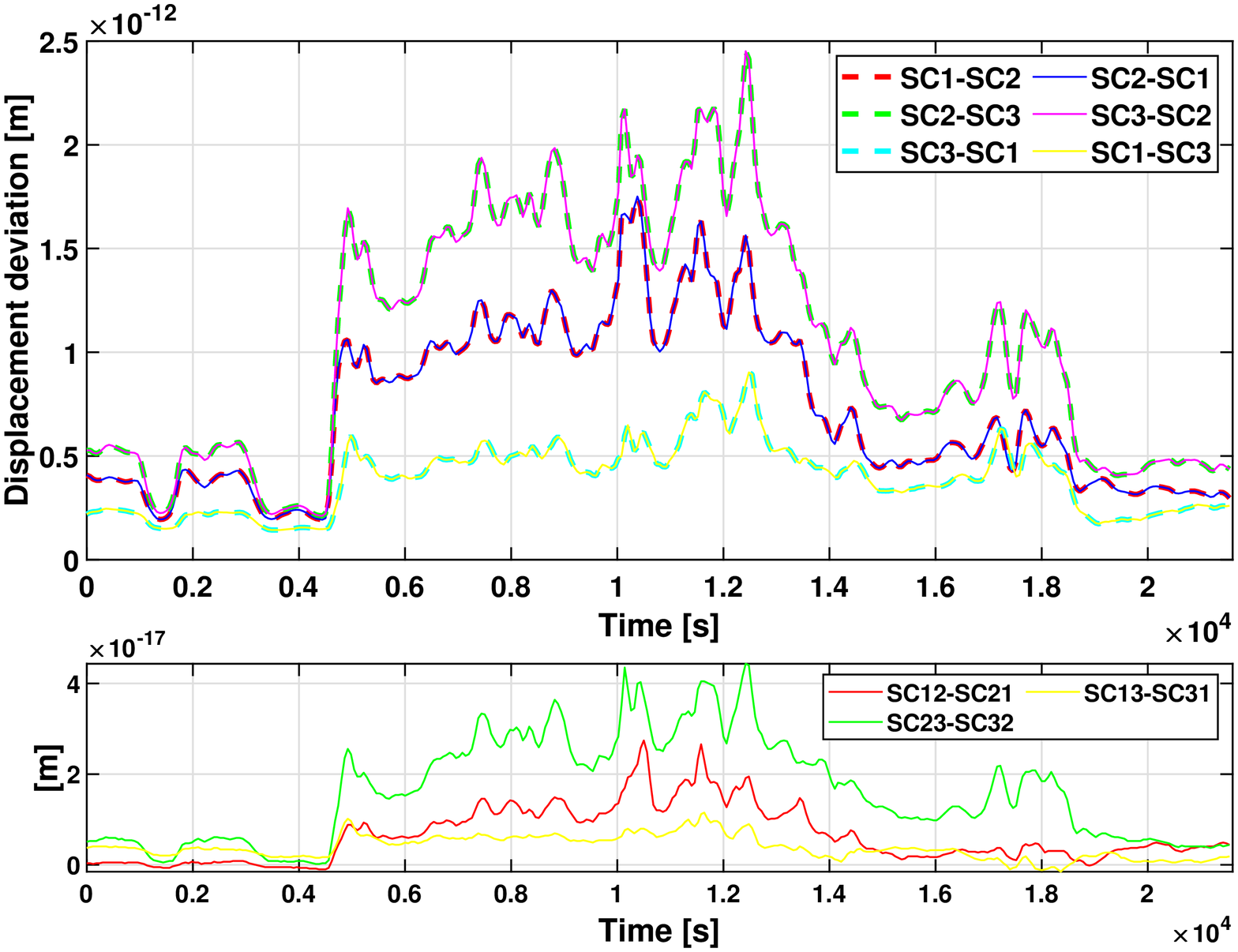}
\caption{\label{6arms_time_60s} Upper plot: Optical path-length variations along the six links with a time resolution of $60~\rm{s}$. The plot shows that the noises along the same arm are almost equal, despite opposite laser propagation directions. Lower plot: The difference between the plasma noises with opposite propagation directions along the same arm. }
\end{figure}

\begin{figure}
\includegraphics[width=0.5\textwidth]{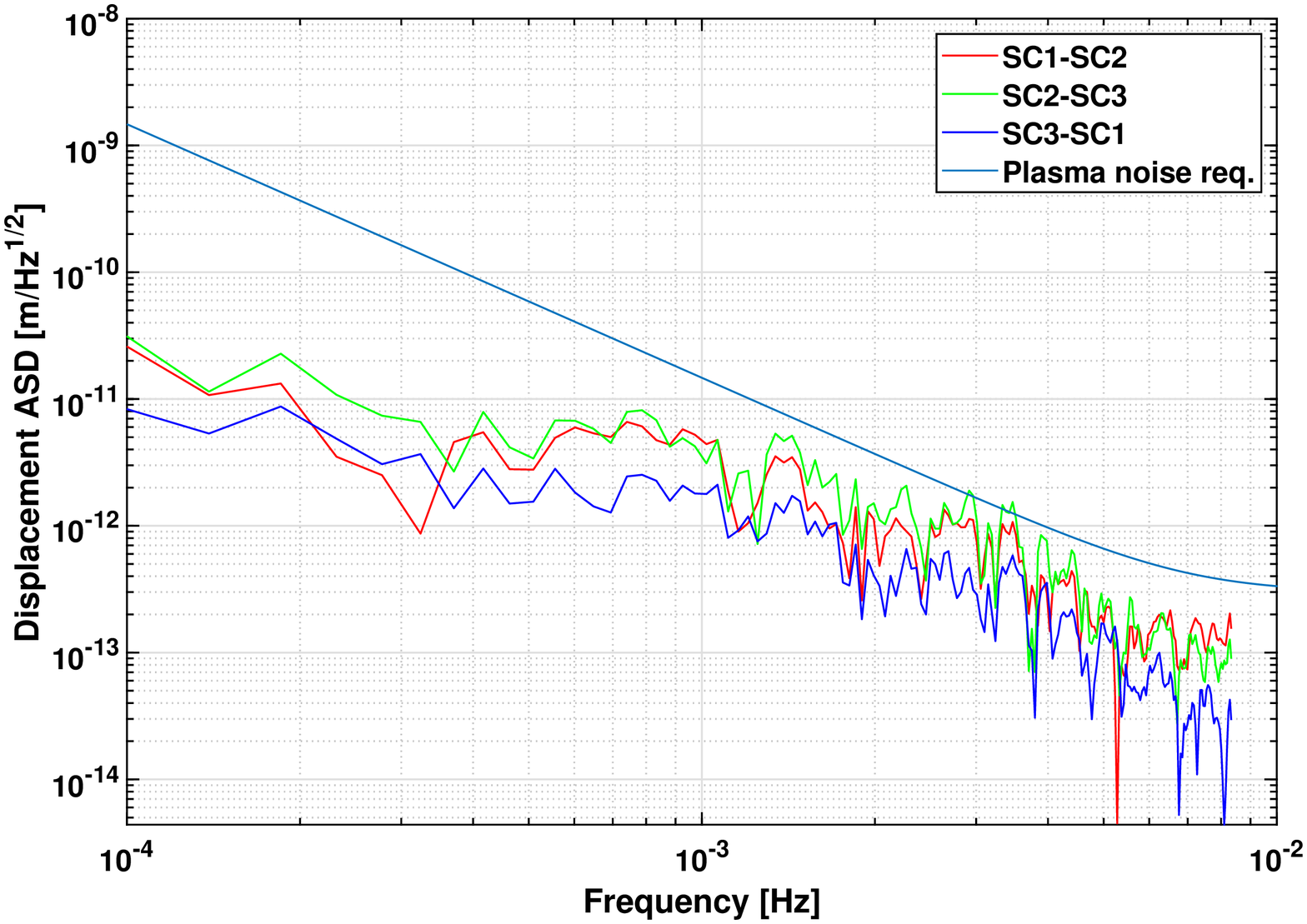}
\caption{The plasma noise ASD curves calculated from the data shown in Fig.~\ref{6arms_time_60s}. The plasma noise requirement is given by Eq. (\ref{eq_req}). } \label{Lu_60_ASD}
\end{figure}

The starting point to properly simulate plasma noise is the generation of real-time plasma density data. Here we use the Space Weather Modeling Framework (SWMF) model, which comes from the NASA Community Coordinated Modeling Center (CCMC), to generate electron number density data with a time resolution of $60~\rm{s}$ \cite{CCMCwebsite}. Taking the real space plasma observation data of the ACE satellite as inputs, we use the SWMF model to obtain the simulation data by solving the magnetohydrodynamics (MHD) equations. By combining the orbital data in Appdendix \ref{NumericalOrbits}, the plasma noise of each link is illustrated in Fig.~\ref{6arms_time_60s}, which includes a magnetic storm ($K_{P}=5$) and in this way gives a noise level higher than normal. Here $K_{P}$ is an index to describe the intensity of magnetic disturbances in near-Earth space \cite{WingKpIndex,LuoKpIndex}. Details on the simulation procedure can be found in Refs.~\cite{LuLF2021raf,Lu2020wavefrontdistortion}. Due to the short light travel time between two satellites ($\sim 0.5$ s), the difference between $s_{ij}(t)$ and $s_{ji}(t)$ is very small (cf. Fig.~\ref{6arms_time_60s}). Eq.~(\ref{signaldelay}) shows that this difference is about $1\times10^{-5}~\rm{pm}$, attributed to the fact that the velocity of TianQin satellites is about $2~\rm{km/s}$. As show in Fig.~\ref{6arms_time_60s}, the dotted and solid lines are overlapped. Therefore we can treat the plasma noises along different propagation directions between a pair of spacecraft as equal. The noise ASD curves of the three arms are shown in Fig.~\ref{Lu_60_ASD}. 

\begin{figure}
\includegraphics[width=0.5\textwidth]{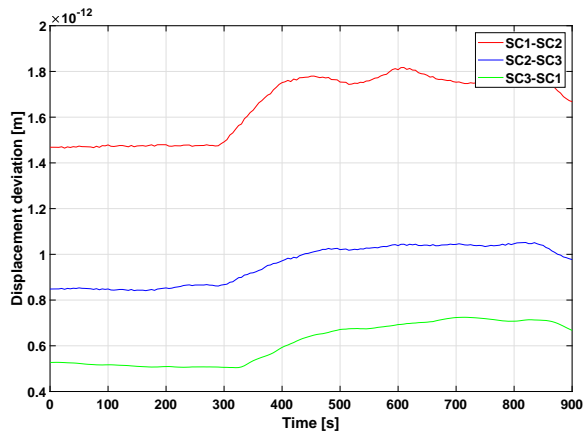}
\caption{Optical path-length variations along the three arms with a time resolution of $5~\rm{s}$.} \label{3arms_time_5s}
\end{figure}

\begin{figure}
\includegraphics[width=0.5\textwidth]{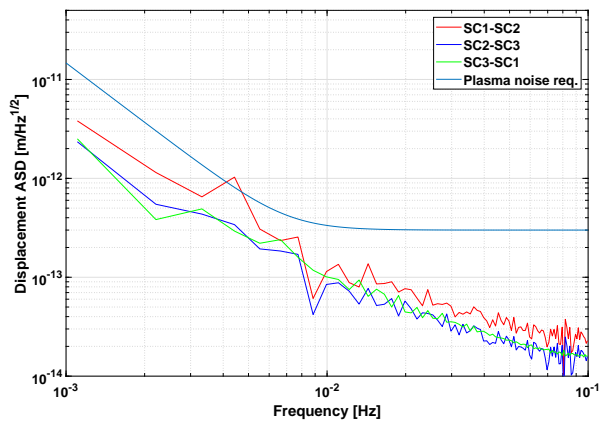}
\caption{The plasma noise ASD curves calculated from the data shown in Fig.~\ref{3arms_time_5s}. The plasma noise requirement is given by Eq. (\ref{eq_req}).} \label{3arms_ASD_5s}
\end{figure}

It is important to emphasize that the spectral density can only reach $8~\rm{mHz}$ due to the limitation of temporal and spatial resolution (the finest spatial resolution of the SWMF model is $0.25~\rm{R_E}$, and $\rm{R_E}$ stands for the Earth radius).  However, for TianQin, we are interested in the behaviour of plasma noise in the range of $10^{-4}-1~\rm{Hz}$. Therefore, in this study we further generate data with a time resolution of $5~\rm{s}$ using the PPMLR-MHD model developed by Hu et al. \cite{HuYQ2005CPL}. Adopting the same methodology as in Ref.~\cite{LuLF2021raf}, we have obtained the results shown in Fig.~\ref{3arms_time_5s}. The ASD of the plasma noise of the three arms is shown in Fig.~\ref{3arms_ASD_5s}. The PPMLR-MHD model has the finest spatial resolution of 0.1 Earth radius in the near Earth region of $-10~\rm{R_{E}}\leq x,y,z\leq10~\rm{R_{E}}$ in the GSE (Geocentric solar ecliptic) coordinate system, and this model has been widely used to study solar wind-magnetosphere-ionosphere system related phenomena \cite{SunTR2020,SunTR2021,FengXS2019}. Note that, unfortunately, due to high cost of such MHD simulations, the 5 s step size and the 900 s data length set the current limits that are available to us. Nevertheless, we deem it sufficient to demonstrate a general trend in the plasma noise spectra. 

Overall, in time domain, we can see that the plasma-induced optical path-length variation along a single link can reach 1 pm level. In addition, as shown in Fig.~\ref{Lu_60_ASD} and Fig.~\ref{3arms_ASD_5s}, we notice that the ASD of the plasma noise decreases with increasing Fourier frequency. In the following subsection \ref{KolmogorovSpectrum}, we show this conclusion is valid in general for plasma noise in the millihertz band. Moreover, the patterns in the Fig.~\ref{6arms_time_60s} and the Fig.~\ref{3arms_time_5s} suggest correlation among the plasma noises along different arms. Indeed, the Table \ref{tab:corr1} below gives the correlation coefficients calculated for the two figures, and the numerical values indicate significant positive correlation among different arms.

\begin{table}[ht]
\caption{\label{tab:corr1} The correlation coefficients calculated from the plasma noise data of the Fig.~\ref{6arms_time_60s} (SWMF model) and the Fig.~\ref{3arms_time_5s} (PPMLR-MHD model). Here SC12 stands for the arm between the satellites SC1 and SC2, and likewise for SC23 and SC13.}
\begin{ruledtabular}
\begin{tabular}{cccc}
& SC12-SC23 & SC12-SC13  & SC23-SC13 \\ 
\hline
SWMF  & 0.9786 & 0.8760 & 0.9348 \\
PPMLR & 0.9828 & 0.9234 & 0.9695 \\
\end{tabular}
\end{ruledtabular}
\end{table}

\subsection{Kolmogorov spectrum}\label{KolmogorovSpectrum}

Solar wind plasma is usually in a turbulent state, and it is hard to describe its properties in time and space accurately. However, a theory that is widely accepted, and consistent with observations is Kolmogorov's statistical theory of turbulence \cite{Thorne2017,Schmidt2010}. The theory suggests that large eddies in a turbulent flow constantly split into smaller ones, and transfer energy in the process. Based on Kolmogorov's theory, the three-dimensional spectral density of the refractive index coefficient can be derived as \cite{Schmidt2010}
\begin{equation}\label{Kolmogorov spectrum}
\Phi^{K}_{n}(k)=0.033C^{2}_{n}k^{-11/3},
\end{equation}
where $k$ is the wave number, and $C^{2}_{n}$ is the refractive-index structure parameter. Similarly, the electron number density variation also follows the power spectral density (PSD) with a $-\frac{5}{3}$ spectral index in one dimension \cite{Jennrich2021okh,Smetana2020raf}.

Recently, in Ref.~\cite{Jennrich2021okh}, the influence of solar wind plasma on LISA science measurements is analyzed based on Kolmogorov's theory. The variation of optical path length on a single link caused by electron number density fluctuation has been obtained \cite{Jennrich2021okh}.
The electron density spectrum is given by
\begin{equation}\label{ElectronDensitySpectrum}
S_{\rm{Ne}}(f)=\frac{4\pi}{(2\pi)^{5/3}}\frac{3}{5}P_{0}k^{11/3}_{0}V^{2/3}f^{-5/3},
\end{equation}
where $P_{0}$ and $f$ are the amplitude and frequency of the electron number density fluctuations, and $k_{0}$ the wavenumber corresponding to the amplitude $P_{0}$, and $V$ the velocity of solar wind. This model of electron density fluctuations in Eq.~(\ref{ElectronDensitySpectrum}) has been verified with observation data coming from the WIND/SWE instrument. The results have shown that the spectral density in the $10^{-4}-10^{-2}~\rm{Hz}$ range agrees well with the Kolmogorov power-law spectrum of Eq.~(\ref{Kolmogorov spectrum}). Further results are shown in Fig.~5 of Ref.~\cite{Jennrich2021okh}, which presents the plasma noise of single links, and they all indicate that each of the 572 daily spectra becomes smaller with increasing Fourier frequency in the range of $10^{-4}~\rm{Hz} - 10^{-1}~\rm{Hz}$.

In Ref.~\cite{Celnikier1983}, two different electron number density spectra were observed in the $10^{-3}-1~\rm{Hz}$ range based on real observations from the ISEE~1-2 spacecraft. In the range of $10^{-3}-6\times10^{-2}~\rm{Hz}$, the spectral index is $-1.67\pm0.05$, which satisfies the Kolmogorov spectrum. Above $6\times10^{-2}~\rm{Hz}$, the spectral index is $-0.9\pm0.2$ \cite{Alexandrova2013}. In Ref.~\cite{Roberts2020},  based on the data from the Magnetospheric MultiScale  mission (MMS), a spectral index of $-0.86\pm0.03$ was observed in the range $0.039-0.97~\rm{Hz}$. In addition, there is extensive literature showing that other parameters of solar wind turbulence follow the Kolmogorov spectrum below $0.1~\rm{Hz}$ \cite{Sahraoui2009PRL,Armano2020zyx,Bruno2013}. These observations are consistent with our MHD simulation results in \ref{MHD_simulation}. 

From the above discussion, it can be seen that the spectrum of electron density fluctuations in the solar wind is roughly composed of two different power-law spectra within $10^{-4}-1~\rm{Hz}$. At the MHD scale (also referred to as inertial range), i.e. in the range of $\sim 10^{-4}-10^{-1}~\rm{Hz}$, density fluctuations follow a Kolmogorov spectrum. At the ion scale corresponding to $\sim 10^{-1}-1~\rm{Hz}$, the density spectrum becomes slightly flatter, but still decreases with increasing Fourier frequency.  \cite{Celnikier1983,Alexandrova2013,Roberts2020,Sahraoui2009PRL,Armano2020zyx,Bruno2013}. In Sec.~\ref{result},  we will use these properties of plasma noise, together with MHD simulation data to generate simulated data in the entire observational frequency band of $10^{-4}-1~\rm{Hz}$.


\section{Plasma noise in TDI}\label{ResidualNoise}
\subsection{Analytical estimation}\label{Theoretical}

In the previous section, we have analyzed the plasma noises of single laser links through MHD simulations and turbulence theory, and the results have shown that the effect has a non-negligible contribution to the noise budget of TianQin. However, in current design of space-based GW detectors, it is necessary to combine multiple measurements from different arms to reduce laser frequency noise in data post-processing, i.e. the TDI technique \cite{Shaddock2003dj,Tinto2020fcc,Tinto2018kij,Otto2012dk}. Therefore, we would like to study how the plasma noises of single links are transferred through TDI combinations, and to derive the corresponding analytical expressions for the residual plasma noises. In the following, for the sake of simplicity, we assume that the laser frequency noise, optical bench displacement noise, etc., can be suppressed below the secondary noise level by TDI combinations.

\begin{figure}
\includegraphics[width=0.4\textwidth]{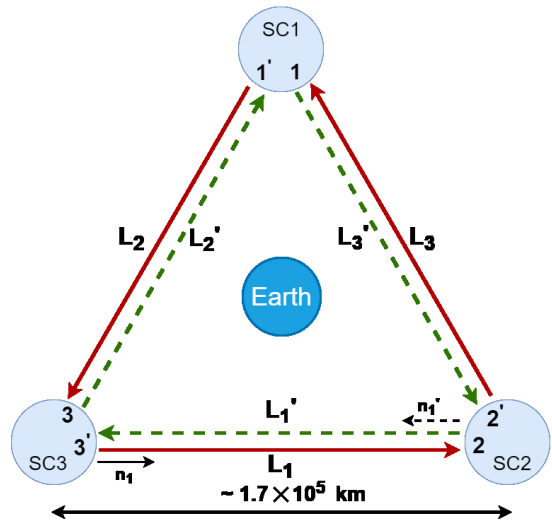}
\caption{Schematic TianQin detector configuration. Three satellites are labeled 1, 2, 3, each with two optical benches, denoted by 1, $1^{\prime}$, 2, $2^{\prime}$, 3, $3^{\prime}$.} \label{TianQin}
\end{figure}

As shown in Fig.~\ref{TianQin}, the TianQin constellation is formed by three identical drag-free spacecraft orbiting around the Earth. Each satellite has two test masses, two optical benches, and two lasers. The distance between each pair of satellites is about $1.7\times10^{5}~\rm{km}$ \cite{Luo2016,Huxc2018}. The following notation is employed, according to the TianQin detector configuration as shown in Fig.~\ref{TianQin}~\cite{Tinto2020fcc,Tinto2018kij,Otto2012dk}. The three spacecraft are marked as S/C $i$, and each satellite carries two identical optical benches ($i$, $i^{\prime}$) connected by optical fibers. The arm lengths of the opposite side of each satellite $i$ are denoted by $L_{i}, L_{i^{\prime}}$, corresponding to the light travel time $t_{i}, t^{\prime}_{i}$, where the unprimed number indicates that the laser beam travels in a clockwise direction, and the primed number a counterclockwise direction. The unit vectors $\vec{\textbf{n}}_{i}$, $\vec{\textbf{n}}_{i^{\prime}}$ are along each arm of the triangular configuration, respectively, and $\vec{\textbf{n}}_{i}=-\vec{\textbf{n}}_{i^{\prime}}$ for each $i$.

The first-generation TDI has different data combinations. We use $\alpha$ and $X$ as examples to calculate the residual plasma noise. The schematic optical paths of the two combinations can be found in, e.g., Ref. \cite{Tinto2020fcc}. Accordingly, the plasma noise in the first-generation TDI $\alpha$ combination can be written as follows \cite{SuWei2021zmj}
\begin{equation}\label{TDIalpha1}
\begin{split}
\delta s_{\alpha1}(t)&=[\mathcal{D}_{1^{\prime}}\mathcal{D}_{2^{\prime}}s_{12}(t)+
\mathcal{D}_{2^{\prime}}s_{23}(t)+s_{31}(t)]
\\&-[\mathcal{D}_{1}\mathcal{D}_{3}s_{13}(t)+\mathcal{D}_{3}s_{32}(t)+s_{21}(t)],
\end{split}
\end{equation}
with the time-delay operators $\mathcal{D}_{i}f(t):=f(t-L_{i}/c)$ and the light speed $c$. 
Similarly, the $X$ combination is given by
\begin{equation}
\begin{split}
\delta s_{X1}(t)&=\mathcal{D}_{3}\mathcal{D}_{2}\mathcal{D}_{2^{\prime}}s_{12}(t)+
\mathcal{D}_{2}\mathcal{D}_{2^{\prime}}s_{21}(t)\\&+\mathcal{D}_{2^{\prime}}s_{13}(t)+s_{31}(t)
-\mathcal{D}_{2^{\prime}}\mathcal{D}_{3^{\prime}}\mathcal{D}_{3}s_{13}(t)
\\&-\mathcal{D}_{3^{\prime}}\mathcal{D}_{3}s_{31}(t)-\mathcal{D}_{3}s_{12}(t)
-s_{21}(t).
\end{split}
\end{equation}

Let us now calculate the residual plasma noise for these combinations\cite{Zhou2021PRD}. First, we transform these expressions from the time domain to the frequency domain through Fourier transform. Then by calculating the expectation value, we obtain the spectral density of the noise. Therefore, the spectral densities of the residual plasma noises corresponding to $\alpha$ and $X$ are given by
\begin{equation}\label{alpha-plasma-noise}
\delta\tilde{S}_{\alpha1}(f)=8\sin^{2}(2\pi fL)S(f),
\end{equation}
\begin{equation}\label{X-plasma-noise}
\delta\tilde{S}_{X1}(f)=\left[16\sin^{2}(2\pi fL)+8\sin(2\pi fL)\sin(4\pi fL)\right]S(f),
\end{equation}
where $S(f)$ is the spectral density of plasma noise of a single link. Here, we assume that the arm length does not change with time and is equal to $L$. Additionally, in the calculations of Eqs.~(\ref{alpha-plasma-noise}) and (\ref{X-plasma-noise}) the following expression is used
\begin{equation}\label{NoiseGeneration}
<s_{ij}(f)\bar{s}_{kl}(f^{\prime})>=\delta(f-f^{\prime})\delta(ij-kl)S(f),
\end{equation} 
where $<>$ denotes expectation values \cite{Cornish2002PRD}. It means that the plasma noises of different arms are uncorrelated. Also recall that the plasma noises of the same arm with different propagation directions are nearly equal (cf. Fig. \ref{6arms_time_60s}), and hereby we neglect the differences. In addition, the other extreme situation corresponds to when the plasma noises of different arms are perfectly correlated with one another. In this case, TDI would cancel the plasma noise in the same manner as it does to the laser frequency noise. 

\begin{figure}
\includegraphics[width=0.5\textwidth]{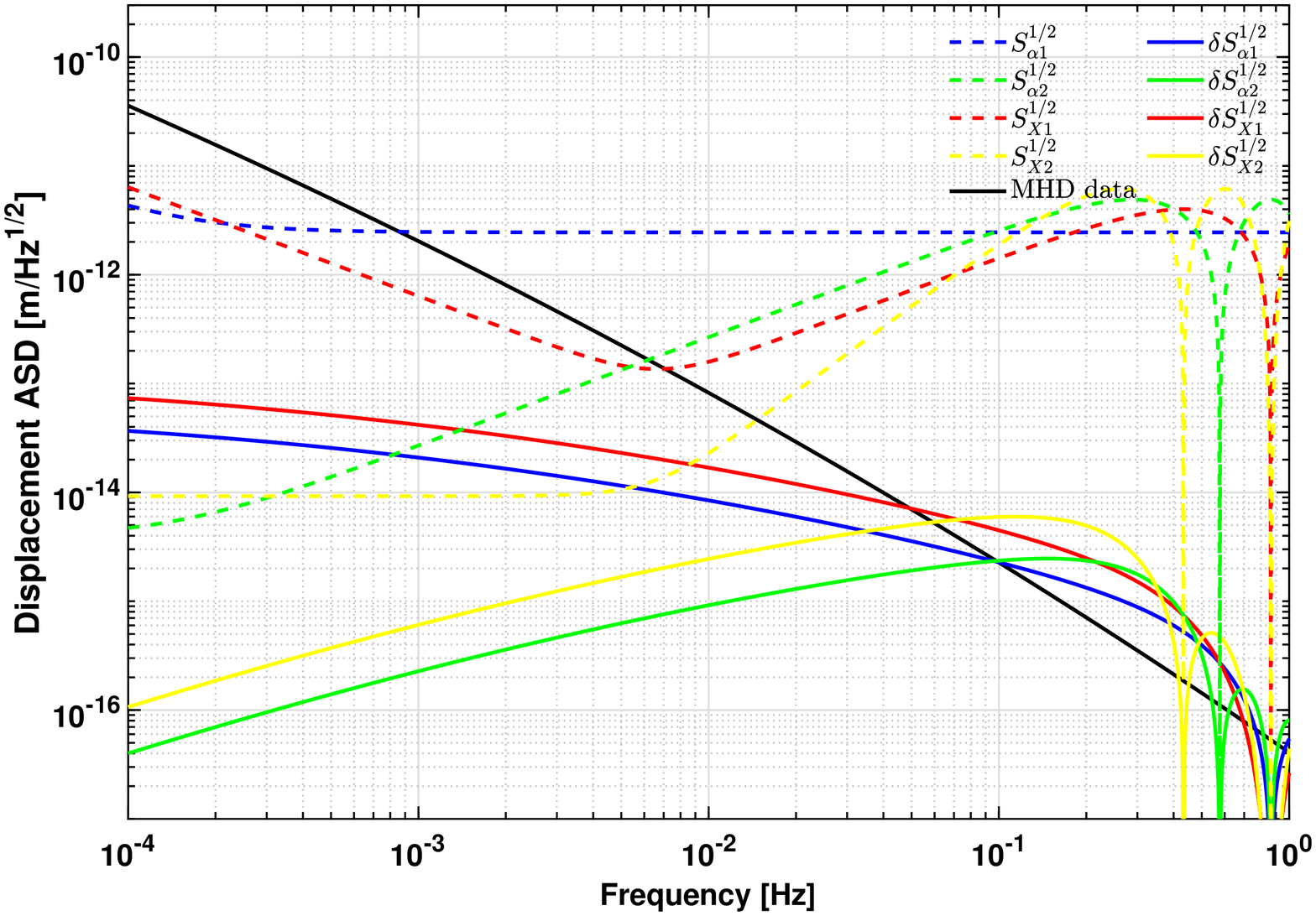}
\caption{\label{residual-noise} Residual plasma noise after the first- and second-generation TDI $X$, $\alpha$ combinations. The solid red and yellow lines represent the plasma noise after the first- and second-generation $\alpha$ combinations, whereas the solid blue and green lines denotes the $X$ combinations. The dashed lines denote the secondary noises (both the acceleration and displacement noises included) after different TDI combinations, with which the solid lines of the same colors should be compared. Additionally, a black line is added to mark the plasma noise of a single link for ease of reference. } 
\end{figure}

The results obtained from Eqs.~(\ref{alpha-plasma-noise}) and (\ref{X-plasma-noise}) are illustrated in Fig.~\ref{residual-noise}. A fitted plasma noise spectral density $S(f)$ is plugged into the residual plasma noise expression, and the former is obtained by extrapolating the spectrum of the MHD data in Fig.~\ref{Lu_60_ASD} to the entire frequency band. Fig.~\ref{residual-noise} shows that below $\sim 0.1~\rm{Hz}$ the TDI-combined plasma noise is smaller than the noise of a single link, while above $\sim 0.1~\rm{Hz}$, the former is larger than the latter. However, since the high-frequency plasma noise itself is quite small, the TDI-combined noise is still smaller than the secondary noise in the frequency band of $10^{-4}-1~\rm{Hz}$. As a comparison,  we also draw the secondary noise for both TDI $\alpha$ and $X$ combinations. The secondary noises are given by the following formulas \cite{Zhou2021PRD,Vallisneri2012CQG,Estabrook2000PRD} 
\begin{equation}\label{alpha1-secondary-noise}
S_{\alpha1}(f)=[4\sin^{2}(3\pi fL)+24\sin^{2}(\pi fL)]S^{\rm{acc}}_{y}+6S^{\rm{opt}}_{y},
\end{equation}
\begin{equation}\label{X1-secondary-noise}
\begin{aligned}
S_{X1}(f)&=[4\sin^{2}(4\pi fL)+32\sin^{2}(2\pi fL)]S^{\rm{acc}}_{y}
\\&+16\sin^{2}(2\pi fL)S^{\rm{opt}}_{y},
\end{aligned}
\end{equation}
in which $S^{\rm{acc}}_{y}=\frac{\left(1\times10^{-15}\rm  ms^{-2}Hz^{-1/2}\right)^{2}}{(2\pi fc)^{2}}=2.8\times10^{-49}(f/1\rm Hz)^{-2}\rm Hz^{-1}$ and $S^{\rm{opt}}_{y}=\frac{\left(1\times10^{-12}\rm mHz^{-1/2}\right)^{2}\times(2\pi f)^{2}}{c^{2}}=4.4\times10^{-40}(f/1\rm Hz)^{2}Hz^{-1}$ are the acceleration and displacement noises of a single link, respectively \cite{Luo2016,Zhou2021PRD}.

Ref.~\cite{Zhou2021PRD} found that the first-generation TDI combinations cannot meet the noise requirement in some frequency bands after considering the numerically optimized orbits. In order to further suppress laser phase noise, TianQin may need second-generation TDI combinations. Therefore, it is desirable for us to analyze how the plasma noise of a single link is transferred in the second-generation TDI combinations to determine whether it affects the overall performance. According to Ref.~\cite{Shaddock2003dj}, the mathematical expressions of the second-generation TDI $\alpha$ and $X$ combinations are given by the following formulas:
\begin{equation}\label{TDI2-alpha}
\begin{split}
\delta s_{\alpha2}(t)&=s_{31}+s_{23;2^{\prime}}+s_{12;1^{\prime}2^{\prime}}+s_{21;3^{\prime}1^{\prime}2^{\prime}}+s_{32;33^{\prime}1^{\prime}2^{\prime}}\\&+s_{13;133^{\prime}1^{\prime}2^{\prime}}-s_{21}-s_{32;3}-s_{13;13}-s_{31;213}\\&-s_{23;2^{\prime}213}-s_{12;1^{\prime}2^{\prime}213},
\end{split}
\end{equation}
\begin{equation}\label{TDI2-X}
\begin{split}
\delta s_{X2}(t)&=s_{31}+s_{13;2^{\prime}}+s_{21;22^{\prime}}+s_{12;322^{\prime}}+s_{21;3^{\prime}322^{\prime}}\\&+s_{12;33^{\prime}322^{\prime}}+s_{31;3^{\prime}33^{\prime}322^{\prime}}+s_{13;2^{\prime}3^{\prime}33^{\prime}322^{\prime}}\\&-s_{21}-s_{12;3}-s_{31;3^{\prime}3}-s_{13;2^{\prime}3^{\prime}3}-s_{31;22^{\prime}3^{\prime}3}\\&-s_{13;2^{\prime}22^{\prime}3^{\prime}3}-s_{21;22^{\prime}22^{\prime}3^{\prime}3}-s_{12;322^{\prime}22^{\prime}3^{\prime}3},
\end{split}
\end{equation}
where the semicolon ';' indicates time delay that is time-dependent, i.e., $f(t)_{;j}=f(t-L_{j}(t)/c)$. 

According to Refs.~\cite{Zhou2021PRD,Krolak2004PRD}, the spectral density of the residual plasma noises for second-generation TDI  $\alpha$ and $X$ combinations are read off as
\begin{equation}\label{alpha2-secondary-noise}
\delta\tilde{S}_{\alpha2}(f)=4\sin^{2}(3\pi fL)\delta \tilde{S}_{\alpha1}(f),
\end{equation}
\begin{equation}\label{X2-secondary-noise}
\delta\tilde{S}_{X2}(f)=4\sin^{2}(4\pi fL)\delta \tilde{S}_{X1}(f).
\end{equation}
Eqs. (\ref{alpha2-secondary-noise}) and (\ref{X2-secondary-noise}) have the same mathematical form as the secondary noises after the second-generation TDI $\alpha$ and $X$ combinations, since we have treated the plasma noise in a similar way to shot noise and test mass acceleration noise. Inserting the expressions (\ref{alpha1-secondary-noise}) and (\ref{X1-secondary-noise}) into the above two equations gives the final residual noise, and the results are shown in Fig.~\ref{residual-noise}. We can see that the residual noise after the second-generation combinations is lower than the secondary noise requirements. These analytic results are confirmed by numerical simulations, which will be described in details in the following subsection. 


\subsection{TDI simulation results}\label{result}

\begin{figure}
\includegraphics[width=0.5\textwidth]{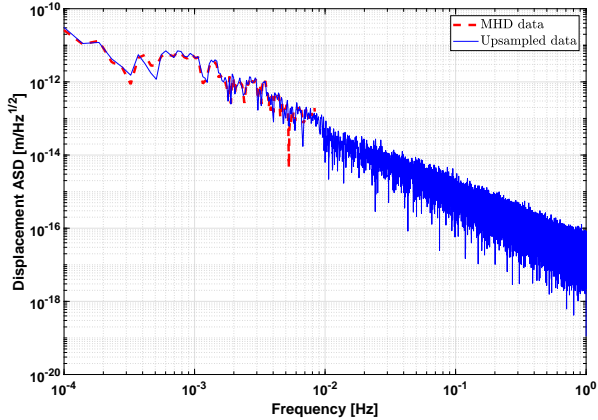}
\caption{Full-band simulation data generated based on Kolmogorov spectrum and the MHD data of Fig.~\ref{6arms_time_60s}.}
\label{dsta combination frequency}
\end{figure}

To verify the analytical studies described above, we further numerically calculate the residual plasma noise after different TDI combinations. We use MHD simulations and the Kolmogorov spectrum introduced in Sec.~\ref{SimulationSetup} to generate simulated data up to $1~\rm{Hz}$ in the frequency band, and the frequency domain results are shown in Fig.~\ref{dsta combination frequency}. The specific procedure is to perform a spectral fit of the MHD simulation data in Fig.~\ref{Lu_60_ASD} and Fig.~\ref{3arms_ASD_5s}, followed by an extrapolation to higher frequencies based on the Kolmogorov spectrum. In this way, we do not alter the trend of the MHD simulation data on a relatively long time scale ($f<8$ mHz), but add some fluctuations that satisfy the Kolmogorov spectrum on a relatively short time scale ($f>8$ mHz). 

\begin{figure}
\includegraphics[width=0.5\textwidth]{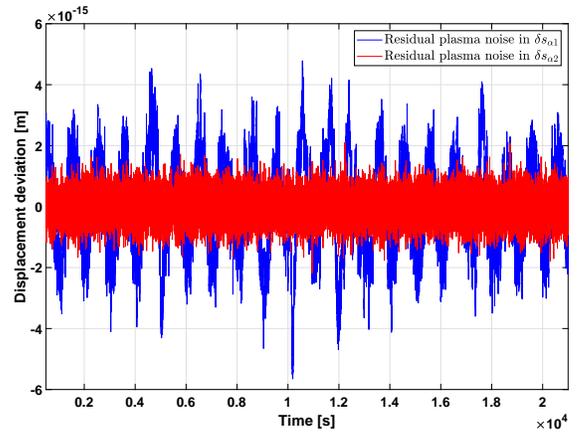}
\caption{Residual plasma noise in time domain for first- and second-generation TDI $\alpha$ combinations.} 
\label{TDI2_60s_time}
\end{figure}

\begin{figure}
\includegraphics[width=0.5\textwidth]{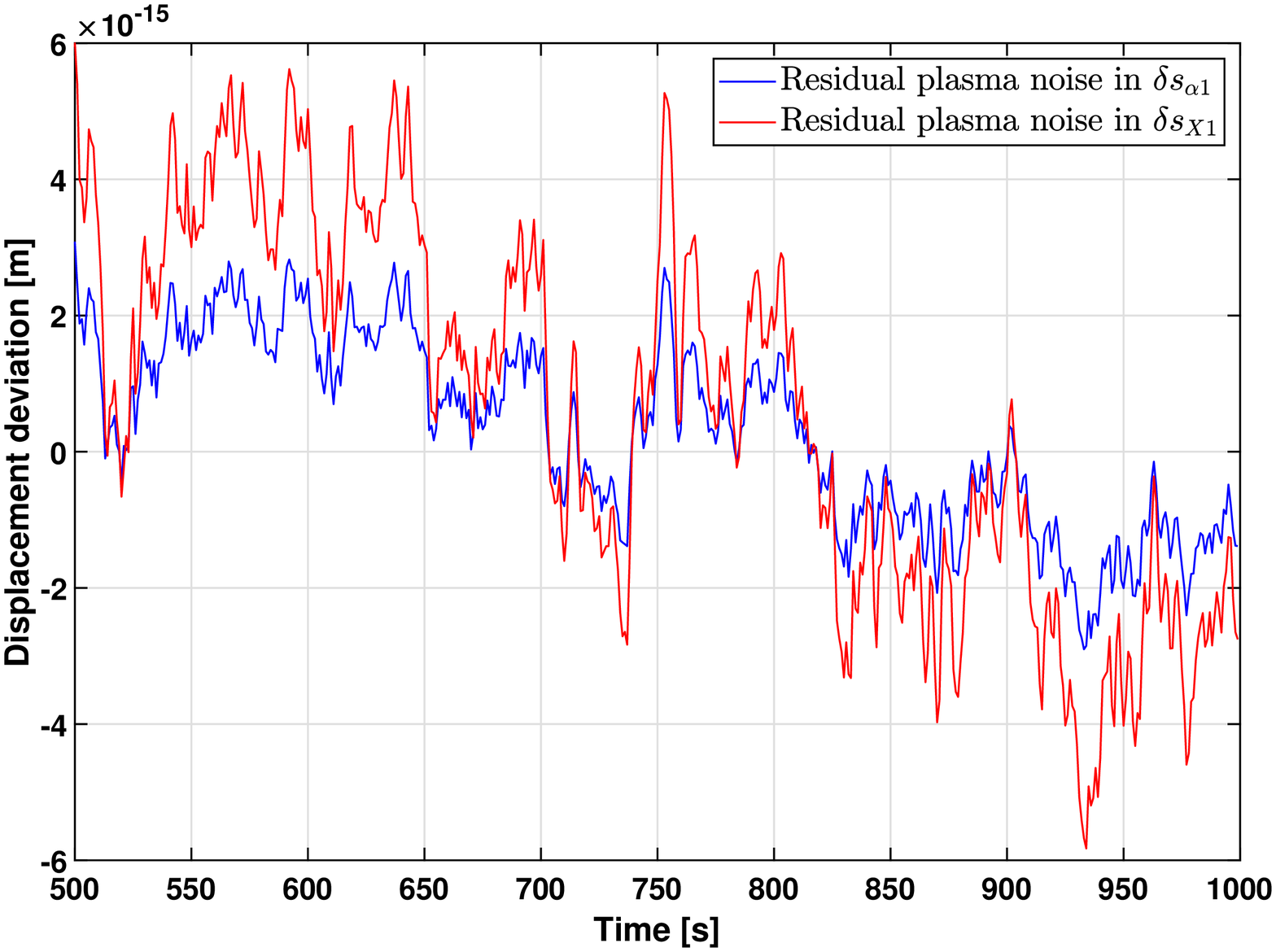}
\caption{Residual plasma noise in time domain for first-generation TDI $\alpha$ and $X$ combinations.} 
\label{TDI_a1_X1_60s_time}
\end{figure}

\begin{figure}
\includegraphics[width=0.5\textwidth]{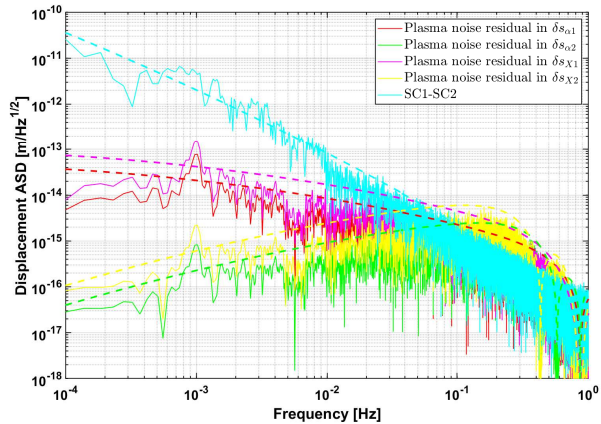}
\caption{ASD of residual plasma noise after first- and second-generation TDI $\alpha$ and $X$ combinations. The corresponding plasma noise of a single arm is shown in Fig.~\ref{6arms_time_60s}. The data above $10^{-2}~\rm{Hz}$ are obtained according to Kolmogorov spectrum by upsampling the MHD data of low frequencies. The dashed line denotes the analytical results. A GW signal with $f_{\rm{GW}}=1$ mHz is also included for testing purposes. } 
\label{TDI2_60s_frequency}
\end{figure}

The time and frequency domain results after different TDI combinations with data at 60 s time resolution (see Fig.~\ref{6arms_time_60s}) are given in Fig.~\ref{TDI2_60s_time}, Fig.~\ref{TDI_a1_X1_60s_time} and Fig.~\ref{TDI2_60s_frequency}, respectively. It can be seen from Fig.~\ref{TDI2_60s_time} that the residual plasma noise after the first- and second-generation TDI $\alpha$ combinations is within $\pm~6\times10^{-15}~\rm{m}$. From Fig.~\ref{TDI_a1_X1_60s_time}, we see that the residual plasma noise after the first-generation TDI $X$ combination is approximately twice as high as the first-generation $\alpha$ combination, which is consistent with our analytical results, and also with the results in Ref.~\cite{SuWei2021zmj}.

Furthermore, the frequency domain results indicate that the above numerical simulation results agree with the analytical results of Sec.~\ref{ResidualNoise}, i.e., the residual noise after the TDI combinations is smaller than the secondary noise in the range of $10^{-4}-1~\rm{Hz}$. For each TDI combination, the residual plasma noise after the TDI combinations is larger than the plasma noise of a single link in the high-frequency band of $10^{-1}-1~\rm{Hz}$. However, since the plasma noise of a single link is relatively small at high frequencies, this amplification will not affect scientific measurements. It should also be remarked that a GW signal with $f_{\rm{GW}}=1~\rm{mHz}$ is included for testing purposes in Fig.~\ref{TDI2_60s_frequency}. 

\begin{figure}
\includegraphics[width=0.5\textwidth]{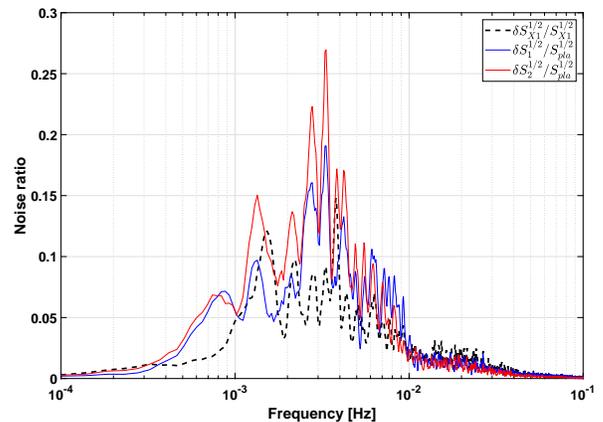}
\caption{The ratios between the plasma noises and the noise requirements for single links (red and blue lines) and TDI $X$ combination (black dashed line).}
\label{Noise_Ratio_TDIX}
\end{figure}

\begin{figure}
\includegraphics[width=0.5\textwidth]{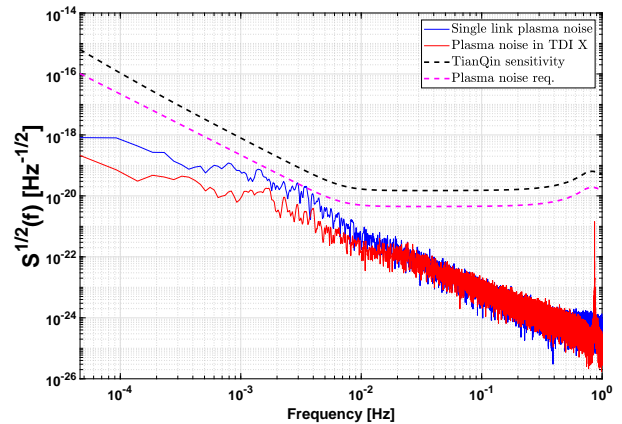}
\caption{Comparison of the plasma noise levels before and after TDI $X1$, and TianQin's sensitivity requirement (TDI $X$). The blue and red lines stand for the plasma noises converted to equivalent GW noises, using the data of the Fig. \ref{TDI2_60s_frequency}. The black and magenta dashed lines represent TianQin's sensitivity goal \cite{Mei2020} and the preliminary plasma noise requirement ($30\%$ of the former, cf. Eq. (\ref{eq_req})), respectively.}
\label{PlasmaSensitivity}
\end{figure}

Fig.~\ref{TDI2_60s_frequency} also shows that the numerical results are slightly smaller than the analytical results after TDI combinations. This is due to the fact that in the analytical formulas we assume no correlation between the plasma noises of different arms. To assess the suppressing effect of TDI for the plasma noise owing to noise correlation, we further calculate the ratio of the plasma noise and the secondary noise requirement before and after TDI $X$ combination, and the results are shown in Fig.~\ref{Noise_Ratio_TDIX}. The results demonstrate that the TDI combination can moderately suppress the plasma noise at the low frequency band below $\sim 10^{-2}~\rm{Hz}$. Another way to see this suppressing effect is to look at the equivalent GW noise levels in the sensitivity diagram (see Fig. \ref{PlasmaSensitivity}), where it shows that the plasma noise level below $\sim 10^{-2}~\rm{Hz}$ indeed becomes lower after TDI combination. 

\begin{figure}
\includegraphics[width=0.5\textwidth]{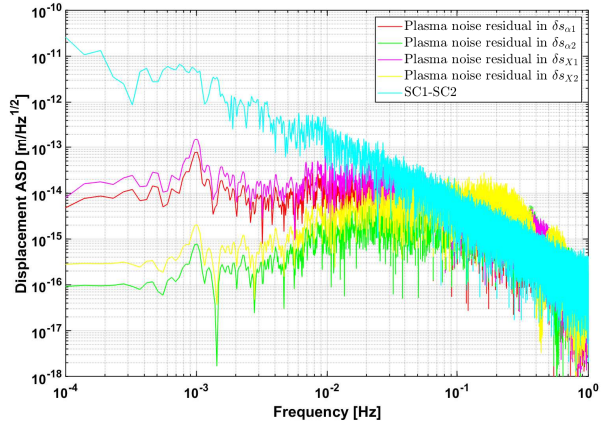}
\caption{ASD after amplifying the high frequency ($f>8~\rm{mHz}$) component of the plasma noise of a single arm in Fig.~\ref{TDI2_60s_frequency} by a factor of 5.} 
\label{TDI2_60s_frequency_times10}
\end{figure}

In the above simulation, we have obtained simulation data over the entire frequency band ($10^{-4}-1~\rm{Hz}$) based on Kolmogorov's turbulent statistical theory by extending the spectrum of the MHD simulation data from $8~\rm{mHz}$ (time resolution $60~\rm{s}$) up to $1~\rm{Hz}$. However, there may be a difference between the high frequency information of the data obtained by this method and the real plasma noise. To be on the safe side, we further multiply the high frequency data in Fig.~\ref{TDI2_60s_frequency} by a factor of 5, and then the results after various TDI combinations are presented in Fig.~\ref{TDI2_60s_frequency_times10}. As it is apparent from the plot, the plasma noise after the combination is still smaller than the secondary noise in the frequency band from $10^{-4}-1~\rm{Hz}$ shown in Fig.~\ref{residual-noise}.

\begin{figure}
\includegraphics[width=0.5\textwidth]{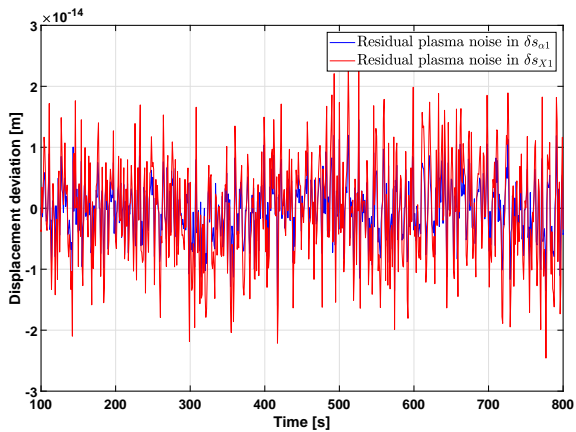}
\caption{Residual plasma noise in time domain for first-generation TDI $\alpha$ and $X$ combinations using the data of Fig.~\ref{3arms_time_5s}.} 
\label{TDIX1_a1_5s_time}
\end{figure}

\begin{figure}
\includegraphics[width=0.5\textwidth]{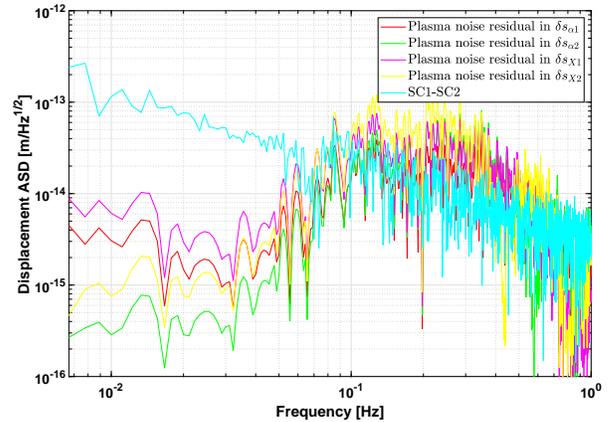}
\caption{ASD of the residual plasma noise after first- and second-generation TDI $\alpha$ and $X$ combinations. The corresponding plasma noise of a single arm is shown in Fig.~\ref{3arms_time_5s}. Up-sampling of the data above $10^{-1}~\rm{Hz}$ is obtained from Kolmogorov spectrum.}
\label{TDI2_5s_frequency}
\end{figure}

To further verify our results above $\sim 10^{-2}$ Hz, we repeat the simulation process using the data from Fig.~\ref{3arms_time_5s}, i.e., using a time resolution of $5~\rm{s}$. The results are displayed in Fig.~\ref{TDIX1_a1_5s_time} and Fig.~\ref{TDI2_5s_frequency}. The residual plasma noise after the first-generation TDI combinations is below $4\times10^{-14}~\rm{m}$, as shown in Fig.~\ref{TDIX1_a1_5s_time}. The results in frequency domain are consistent with the results in Fig.~\ref{TDI2_60s_frequency}.

It should be emphasized that in the second-generation TDI simulations, we have used the following approximation
\begin{equation}\label{LagrangeApprox}
f(t)_{;jk}=\mathcal{D}_{j}\mathcal{D}_{k}f(t)=f\left(t-\frac{L_{j}}{c}-\frac{L_{k}}{c}\right),
\end{equation}
where we have ignored higher order corrections. Considering that the high-frequency component of the plasma noise is quite small, the approximation in Eq.~(\ref{LagrangeApprox}) is reasonable. In the simulations, we have used a 31st-order Lagrange fractional delay filter \cite{Shaddock2004PRD}.


\section{Conclusion}\label{summary}

In this paper, we first obtain two sets of plasma noise data with the time resolutions of 60 s and 5 s using the SWMF model \cite{SuWei2021zmj,LuLF2021raf,Lu2020wavefrontdistortion,Su2020acceleration} and the PPMLR-MHD model \cite{HuYQ2005CPL}, respectively. Then the frequency domain characteristics of the plasma noise along single links are analyzed, and it is shown that the noise behavior can be approximated by a Kolmogorov spectrum in the frequency range of $10^{-4}-1~\rm{Hz}$. According to Fig.~\ref{Lu_60_ASD} and Fig.~\ref{3arms_ASD_5s}, the plasma noise level decreases as the Fourier frequency increases. Especially above $1\times 10^{-2}$ Hz, it becomes negligible compared with the noise requirement. 

Since TDI plays a key role in GW data processing, we further evaluate the plasma noise in TDI combinations both analytically and numerically, and the results agree with each other. The evaluations show that the effect is below the secondary noise except for strong magnetic storms, and hence does not constitute a showstopper to the mission. In addition, we have calculated the ratio of the plasma noise to the secondary noise requirement before and after TDI combinations and found the ratio becomes smaller after the TDI X combination, indicating that TDI has moderate suppressing effect for the plasma noise in terms of GW strain sensitivity (cf. Fig. \ref{PlasmaSensitivity}), due to the correlation of the plasma noises among different arms of TianQin (cf. Fig. \ref{6arms_time_60s}). Our conclusion is consistent with the results in Ref.~\cite{SuWei2021zmj}. 

In the TDI simulations, we have used the Kolmogorov spectrum to extend the fitted spectrum of single-link plasma noises from low frequencies up to high frequencies and obtained the simulated data in the entire measurement band. By cross-checking with the new MHD simulation with the time resolution of $5~\rm{s}$, we deem that the simulation results below $0.1~\rm{Hz}$ are reasonable. Above $0.1~\rm{Hz}$, real solar wind observations show that the plasma noise continues to decrease in amplitude as the Fourier frequency increases \cite{Alexandrova2013,Celnikier1983}. In addition, a minor magnetic storm ($K_{P}=5$) is included in all of the above MHD simulation data. When the solar activity is at extreme levels, the plasma noise after TDI combinations may become greater than the secondary noise primarily near $4~\rm{mHz}$. In this event the data quality is degraded, and extra attention would be needed in GW data processing on ground. 

\begin{acknowledgments}
This work was carried out using the SWMF and BATS-R-US tools developed at the University of Michigan's Center for Space Environment Modeling (CSEM). The modeling tools are available online through the University of Michigan for download and are available for use at the Community Coordinated Modeling Center (CCMC). We thank Hsien-Chi Yeh and Jun Luo for helpful discussions. Special thanks to the anonymous referee for valuable suggestions on improvement. X.Z. is supported by the National Key R\&D Program of China (No. 2020YFC2201200). L.L. is supported by NSFC Grant No. 42004156. The TDI simulations performed in this study have utilized the TQTDI program developed by L.Z., S.Y, and X.Z.
\end{acknowledgments}

\appendix
\section{Numerical orbits}\label{NumericalOrbits}

In addition to requiring the variation of electron number density in space with time, we also need the orbital data of three satellites. Here, we use NASA developed open source software General Mission Analysis Tool (GMAT) to generate orbital data with $1~\rm{s}$ step size \cite{GMATsoftware}. This software has been previously used to study the problems associated with the orbit of TianQin \cite{YeBB2019,TanZB2020,ZhangXF2020PRD,YeBB2020Eclipse}. The starting time of the satellite orbit is 10 Aug 2034 00:00 UTC, which is within the detection period of 3+3 month observation window \cite{Luo2016}. To maintain compatibility with the MHD simulation data in the previous subsection, we adopt the Geocentric Solar Magnetospheric (GSM) coordinates. In this coordinate system, the origin at the center of the Earth has its \emph{x}-axis points to the Sun. The \emph{y}-axis is perpendicular to the Earth's magnetic dipole axis. The positive \emph{z}-axis is aligned with the Earth's northern magnetic dipole axis.

On the other hand, we use the spline interpolation method to obtain the inter-satellite light propagation time and the satellite's position at the moment the light is emitted and received.

\section{Comparing noise correlation with LISA} \label{appdx:NoiseCorrelation}

Our demonstration of significant plasma noise correlation between different arms for TianQin may initially appear deviating from LISA's study in \cite{Jennrich2021okh}, where it claims that such correlation is negligible for LISA based on the analysis in the Appendix A2 therein. To help understanding this apparent difference, here we provide the following possible explanations. 

First, the more precise statement of LISA's recent estimation (\cite{Jennrich2021okh}, also cf. Eq. (A24)) is that the magnitude of the cross-correlation terms (i.e. between arms) is smaller than the auto-correlation terms (i.e. single arms) ``by a factor of order $\omega_0⁄\omega$''. Here one has the key parameter $\omega_0 = 2\pi V/L$ with $V$ the velocity of the solar wind ($\sim 400$ km/s) and $L$ the arm-length. For LISA, the threshold cycle frequency is given by $f_0 = \frac{\omega_0}{2\pi} = \frac{400 \mathrm{km/s}}{2.5\times 10^6 \mathrm{km}} = 1.6\times 10^{-4} \mathrm{Hz}$, above which the cross-correlation is neglected in \cite{Jennrich2021okh}. This frequency is quite low and may have covered the frequency region ``of most interest'' for LISA. But for TianQin, if \emph{ignoring} the Earth, one has $f_0 = \frac{400 \mathrm{km/s}}{1.7\times 10^5 \mathrm{km}} = 2.4\times 10^{-3} \mathrm{Hz}$, which overlaps much more with TianQin's measurement band of 0.1 mHz -- 1 Hz due to the \emph{shorter} arm-length. Hence, this indicates that one may not neglect the cross-correlation for TianQin. 

Second, global MHD simulations allow one to directly evaluate the noise correlation for TianQin from electron-density distribution. But unfortunately, non-local observation/simulation data might not have been available for \cite{Jennrich2021okh}. Instead, analytical modeling relying mostly on the Taylor hypothesis, and \emph{in situ} single-point measurement data were used. 

Third, from the overall physical picture, because the electron-density fluctuations near the detectors are originated from the same source, i.e., the solar activities (e.g. solar storms), and, in TianQin's case, affected by the same Earth's magnetic field, one may expect at least some level of correlation among the three arms, though the level may vary at different frequencies and dynamical scales. Further modeling of plasma noise correlation is worth investigating in future works.

\end{document}